\documentclass[10pt, twocolumn]{svjour3}         
\smartqed  
\usepackage{graphicx}
 \usepackage{mathptmx}      
\usepackage{latexsym}
\usepackage[
authoryear,
round,
sort]{natbib}
\begin{document}

\title{Characterization of the Reactive  Flow Field Dynamics in a Gas Turbine Injector Using High Frequency PIV}


\author{S. Barbosa        \and
        Ph. Scouflaire \and
        S. Ducruix
}


\institute{S. Barbosa, Ph. Scouflaire, S. Ducruix \at
              Laboratoire EM2C - CNRS \\
              Ecole Centrale Paris\\
              Grande Voie des Vignes\\
              92295 Ch\^atenay-Malabry, France\\
              Tel.: +(33) 1 41 13 10 86\\
              Fax: +(33) 1 47 02 80 35\\
              \email{severine.barbosa@em2c.ecp.fr}           
}

\date{Received: date / Accepted: date}

\maketitle

\begin{abstract}
The present work details the analysis of the aerodynamics of an experimental swirl stabilized burner representative of gas turbine combustors. This analysis is carried out using High Frequency PIV (HFPIV) measurements in a reactive situation. While this information is usually available at a rather low rate, temporally resolved PIV measurements are necessary to better understand highly turbulent swirled flows, which are unsteady by nature. Thanks to recent technical improvements, a PIV system working at 12 kHz has been developed to study this experimental combustor flow field. Statistical quantities of the burner are first obtained and analyzed, and the measurement quality is checked, then a temporal analysis of the velocity field is carried out, indicating that large coherent structures periodically appear in the combustion chamber. The frequency of these structures is very close to the quarter wave mode of the chamber, giving a possible explanation for combustion instability coupling. The frequency of these structures is very close to the quarter wave mode of the chamber, giving a possible explanation for combustion instabilities coupling.
\end{abstract}

\section{Introduction}
\label{intro}
Stronger regulations concerning pollutant emissions led to the development of new generations of swirl burners stabilizing lean premixed flames well suited to reach low NOx levels \citep {Correa:1998}.
Swirling motion permits to improve the mixing rate between fuel and oxidant streams and to control the flame stabilization through the swirl-induced recirculation of hot products near the nozzle \citep{Chen:1988, Syred:2006}. Above a critical swirl number ($S > 0.6$), the expansion of the swirling flow leads to a vortex breakdown, which is accompanied by an inner recirculation area. This zone provides the major mechanism for flame stabilization since this recirculation area of hot combustion products continuously supplies energy for the ignition of the incoming fuel-air mixture.\\  
However, under lean conditions, these flames tend to exhibit undesired combustion instabilities, which may cause flame extinction, flashback, reduce engine life time or lead to catastrophic structural damages \citep{Candel:2002, Nauert:2007}, despite the stabilizing effect of their recirculation zones. The mechanisms of these instabilities are based on various complex interactions between combustor geometry, flow field, pressure, mixing, chemical reactions and heat release. It has been shown that swirled flows develop characteristic periodic large coherent structures, which play an essential role in the dynamics of turbulent swirled flames \citep{Paschereit:1998}. Their interactions with the heat release process and acoustic resonant modes of the combustor can cause undesirable thermo-acoustic instabilities in the combustion system.
Due to the importance of these structures as drivers of combustion instabilities, numerical and experimental studies have been undertaken in order to understand their interactions with chemistry and acoustics \citep{Roux:2004, Sengissen:2007:Lar}.\\
Lab scale swirl stabilized combustors have been developed and studied using microphones, $OH^*$ or $CH^*$ emission sensors \citep{Paschereit:1999}, photographic imaging technique \citep{Broda:1998} or laser diagnostics \citep{Weigand:2005, Janus:2005, Weigand:2006, Meier:2006, Olivani2007}.\\ 
Laser-based tools offer the potential to measure most of the important quantities necessary to gain a deeper insight into complex chemical and physical processes in turbulent flames. The flow field can be measured by Laser Doppler Velocimetry (LDV) or Particle Imaging Velocimetry (PIV) \citep{Mueller:1998, Ji:2002, Woodmansee:2007}, mixing process and flame structure by Planar Laser-Induced Fluorescence (PLIF), while laser Raman scattering permits to determine the major species concentrations, temperature, and mixture fraction. Up to now, due to laser technology limitations, these information were usually sampled with a repetition rate of an average of 10 Hz (except for LDV), giving combustion statistical quantities that significantly helped in understanding turbulent flames. However, some of the mechanisms leading to combustion instabilities are still not well understood, as the interactions existing between large coherent structures, chemical reactions and the acoustics of the burner are complex and usually strongly dependent on the configuration.
Indeed the number of experimental studies investigating the temporal evolution of the flow field in gas turbine combustion chambers is rather limited. This is partly due to technological limitations of lasers and cameras, which do not permit to acquire flow field data at a high acquisition rate. However, combustion instabilities, flame extinctions or flashbacks are by nature very unsteady phenomena, whose comprehension is built on the study of spatially and temporally resolved data.\\
In the present work, the flow field of a lean swirl stabilized flame in a gas turbine model combustor is investigated using time resolved PIV. For the last 20 years, the development and improvement of PIV have proceeded in several stages, related to the technical progresses successively achieved in the fields of lasers (double-pulsed solid-state lasers), cameras (now holding two images recorded in rapid succession) and computers (for post-processing) \citep{adrian:2004}. The standard system now usually comprises a single fast camera, a double pulsed Nd:Yag laser and a fast hard-wired PIV correlator.\\ 
The basic method used to extract the velocity field from two successively acquired pictures can be decomposed in several steps. First, the raw image pairs are divided into several windows, called interrogations areas. Then, each pair of corresponding windows are {cross-correlated} using Fast Fourier Transform algorithms \citep{Scarano:1999}. A high cross-correlation value is observed where many particles match up with their corresponding spatially shifted partners, and small cross-correlation peaks may be observed when individual particles match up with other particles. Only the highest correlation peak is interesting as its position in the correlation plane directly corresponds to the average particle displacement within the interrogation area investigated. Finally, the velocity can be determinated, since the time separation between the two pictures of one image pair is known. It appears that many parameters influence the velocity measurement and calculation: the number of particles image pairs in the interrogation area \citep{Adrian:1991}, the size of the particles with respect to their spatial discretisation and the peak interpolation scheme. Recently, many efforts have been made to develop and improve more robust PIV algorithms: the spatial resolution and the measurement precision have been increased. The primary step was to reduce the size of the interrogation area from $64\times64$ pixels or larger \citep{Adrian:1991} to less than $10\times10$ pixels \citep{Nogueira:2001} and to use window offset to increase the spatial resolution \citep{Westerweel:1997}. A secondary step was to reduce the uncertainty associated with the interrogation procedure by using iterative methods \citep{Scarano:1999}, for example. As velocity measurements are necessary and now widely used to study turbulent flows, improvements are continuously carried out to increase the spatial resolution \citep{Lavoie:2007} or to reduce statistical errors (see \citet{Nogueira:2001, Angele:2005, Cholemari:2007} among others).\\ 
However, if a higher spatial resolution can be needed to get a finer comprehension of turbulent flows, a higher time resolution is also strongly necessary. For a long time, due to technical limitations, it was not possible to carry out high frequency PIV measurements (HFPIV). But recent improvements in the performances of computers, lasers and cameras make this development now possible. Recently, PIV systems operating at an acquisition rate up to 20~kHz have been used to study the wake in bluff bodies \citep{Williams:2003} or to obtain velocity fields for both cold and hot flows \citep{Wernet:2007}. However the PIV system used by \citet{Williams:2003} consists in a copper vapour laser and a dump camera. This induces a heavy post-processing work and seems less adapted to the acquisition and the analysis of large series of velocity fields.\\
In the present work, the flow field of a swirl-stabilized combustor, representative of Gas-Turbine combustor, is studied with a HFPIV system at an acquisition rate of 12~kHz. The structure of this paper is as follows. In section~\ref{sec:1}, the experimental burner and the PIV system are described. Then, the structures of the mean reactive flow are detailed in section~\ref{sec:55}, so that the behavior of the burner can be studied and the efficiency and the resolution of the new system can be tested. Finally, instantaneous flow fields are presented and their analysis permits to gain deeper insight into the dynamical interactions between the flame, the vortices and the turbulent surrounding flow in the combustion chamber.

\section{Experimental Configuration} \label{sec:1}
\subsection{Experimental Configuration} \label{sec:2}
\begin{figure}[!h] 
\centering
\includegraphics[width= 5cm]{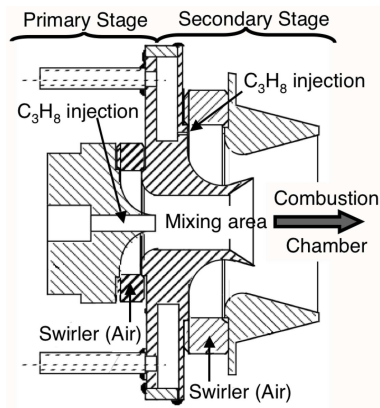} 
\caption{Schematic view of the experimental injector.}
\label{fig:1}
\end{figure}

Figure~\ref{fig:1} shows a schematic diagram of the experimental injector. The setup is composed of a two-staged swirled injector, representative of a gas turbine injector,  supplied with propane and air and a rectangular combustion chamber.
\\
The primary stage consists in a central duct fed with pure fuel and a swirler supplied with air. The angle of the primary stage swirler, which has 18 vanes, is maintained at 42\textsuperscript{o}. The fuel/air mixture is realized in a mixing area downstream the injections of propane and air. Fuel is delivered in the secondary stage by 15 holes. These 15 jets are located on a circular hollow part fed with propane. The airflow of the secondary stage is injected through a swirler with 20 vanes. The angle of the secondary stage swirler is 35\textsuperscript{o}. Both air and combustible are mixed together in the secondary stage. Both swirlers turn the flow in the same direction, in order to obtain a strong co-rotating swirl motion. The primary and secondary fuel/air mixtures mix together while entering the combustion chamber.\\
Both air swirlers are permanently supplied with air and the airflow rate in the secondary stage is 4 times higher than the first stage one. For the present study, for ease of illustration, only the secondary stage fuel injection is used.
Dried compressed air is available at a pressure of 0.7 MPa, while propane is stored at 0.5 MPa. Air and propane mass flows are monitored with electronic mass flow meters and controllers (\textit {Bronkhorst-Elflow}).\\
The combustion chamber has a square cross section of $100 \times 100$ $\textrm{mm}^{2}$ and a length of 500 mm. The side walls of the chamber are made of two silica windows in order to enable optical diagnostics in the flame, while the top and bottom walls are made of concrete. All tests are performed at ambient pressure and ambient temperature.

\subsection{HFPIV measurements} \label{sec:3}
\begin{figure}[!h] 
\centering
\includegraphics[width=8.4cm]{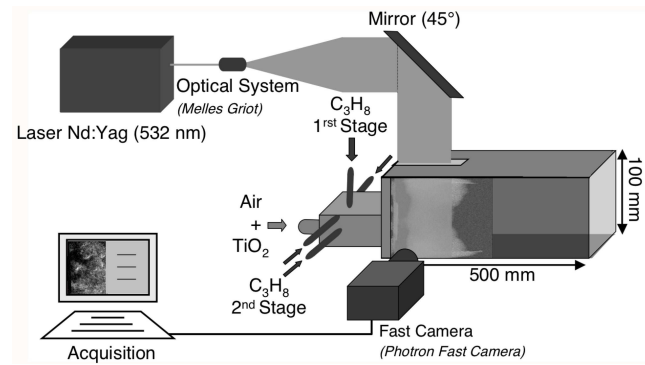} 
\caption{Schematic view of the experimental combustor and the PIV system.}
\label{fig:2}
\end{figure}

\begin{figure}[!h] 
\centering
\includegraphics[width=5cm]{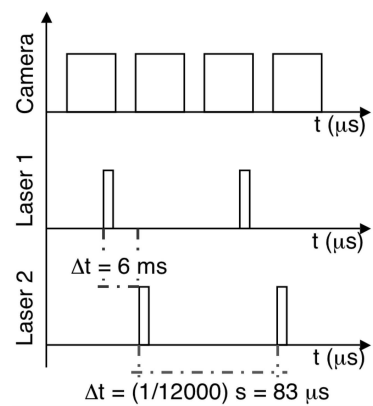} 
\caption{Time diagram for the time resolved PIV.}
\label{fig:3}
\end{figure}

A High Frequency Particle Image Velocimetry (HFPIV) system is used to measure velocity flow fields in the combustion chamber. A schematic view of the experimental burner and the PIV system are reported on figure~\ref{fig:2}.
\\
For such experiments, the top and bottom walls of the chamber contain rectangular silica windows (25~mm wide and 120~mm long) that allow the laser sheet to cross vertically the combustion chamber, downstream the injection plane. 
The light sheet is generated by a system consisting of two Nd:YAG lasers (\textit{Quantronix}), a laser beam recombining device and a set of cylindrical lenses (\textit{Melles Griot}). Both lasers  emit a pulse  with a wavelength of 532~nm. They have a pulse energy and temporal width of 6~mJ and 160~ns respectively. Optics are used to combine both beams along the same trajectory. A set of lenses is used to transform the laser beam into a planar light sheet 60~mm wide and 0.3~mm thick. A fast speed camera (Photron Fastcam, $1024\times1024$ pixels at a rate of 2000 frames per second) equipped with a 105 mm F/1.8 Nikon Nikkor objective is placed perpendicularly to the burner axis to record the image pairs.\\
The two lasers working at 12~kHz and the camera operating at 24~kHz are synchronised by a pulse delay generator (\textit{BNC} 555 pulses/delay Generator), as shown in the time diagram of figure~\ref{fig:3}. 
\\
At this rate of 24~kHz (for the camera), the size of the pictures is $128\times512$ pixels, a limited size which will be improved using new generation cameras. The airflow rate is seeded with TiO2 particles with a nominal diameter $d = 1 \, \mu$m. TiO2 has been used because its melting point is higher than the adiabatic flame temperature of the flame. These seeding particles will be present both in fresh and burnt gases. Using estimations of the maximum velocity, the time delay between the two pulses is chosen equal to $\Delta t=6 \, \mu$s.\\
The raw image pairs are then exported and an off-line image processing is performed with an adaptative {cross-correlation} program ("Flow-Manager" by \textit{Dantec}) using Fast Fourier Transform algorithm. The raw image pairs are divided into squared interrogation areas whose final dimensions are $8\times8$ pixels with an overlap of 25 \%. Finally, the spatial resolution of the flow field measured is $1.07\times1.07$~ mm$^2$. The PIV program is designed and optimized to deal with high velocity and density gradients, providing a peak finding error of less than 0.1~pixel in classical applications \citep{Westerweel:1997}.
The parameters used for the HFPIV treatments are indicated in table~\ref{tab:1}.
\begin{table}[!h]
\centering
\caption{PIV acquisition and processing parameters. All the images have the same size, however the region observed in the combustion chamber can be modified as described in section~\ref{sec:5}. $D$ is the outer diameter of the injector.}
\label{tab:1}       
\begin{tabular}{ll}
\hline\noalign{\smallskip}
{Acquisition} & {} \\
\noalign{\smallskip}\hline\noalign{\smallskip}
{Field-of-view}&{$x=0.47D; y=1.6D$} \\
{Image size}&[128 512] pixels \\ 
{Camera pitch}& 0.017 mm/px\\ 
{Frequency}& 12 kHz\\  
{Nb of raw images}& 24,000\\
\hline\noalign{\smallskip}
{Processing} & {}\\
\noalign{\smallskip}\hline\noalign{\smallskip}
{Sub-pixel scheme}&Gaussian \\
{Initial window size}&[32, 32] pixels \\ 
{Final window size}&[8, 8] pixels \\ 
{Overlap}& 25\% \\  
{Nb of vectors}& [21, 85] \\
\noalign{\smallskip}\hline
\end{tabular}
\end{table}
\subsection{Post-processing} \label{sec:4}
For each experiment, a continuous acquisition of $N = 12 000$ image pairs ($24 000$ raw images) corresponding to a physical time of 1~second has been performed. The instantaneous velocity fields associated with the $N$ image pairs are processed to determine the mean velocity components $U_{mean}$ and $V_{mean}$ and the RMS velocity components $U_{rms}$ and $V_{rms}$ (for "root mean square"). In the present study, the notation $u$ and $v$ refer respectively to the horizontal and vertical velocity components.
\begin{equation}
U_{mean} = \frac{1}{N} \sum_{i=1}^{N} u_i
\end{equation}
\begin{equation}
U_{rms} =\sqrt{ \frac{1}{N} \sum_{i=1}^{N}( u_{i}-U_{mean})^{2}}
\end{equation}
Each instantaneous velocity field is also processed in order to obtain the instantaneous vorticity field in the plane $(\vec{x},\vec{y})$ using the following definition of the local vorticity, $\omega$:
\begin{equation}
\omega = \frac{\partial{u}}{\partial{y}} -\frac{\partial{v}}{\partial{x}}
\end{equation}
where $\vec{x}$ refers to the horizontal direction and $\bf{y}$ to the vertical direction. As mentioned before, the spatial resolution of the velocity fields measured is rather low, making the calculation of $ \partial{u}/ \partial{y}$ and $\partial{v}/\partial{x}$ difficult. Nevertheless,  as it will be discussed in section~\ref{sec:8}, this grid is fine enough to obtain interesting results concerning the behavior of the experimental burner.\\
While statistical quantities are necessary to get a good comprehension of the burner behavior, time resolved results are also necessary to improve the understanding of the strongly unsteady phenomena occuring in the combustor. As mentioned in the introduction, these HFPIV measurements are mainly  carried out to investigate the development of coherent structures in the combustion chamber. As, a clear and universal definition of coherent strutures does not exist, \citet{Jeong:1995} consider the following to be the requirements for a vortex core:
\begin{itemize}
\item a center of a vortex corresponds to a local pressure minimum
\item a vortex core must have a net vorticity
\item the geometry of the identified vortex core should be Galilean invariant.
\end{itemize}
Unfortunately, these three requirements do not permit to build a single identification scheme that can be used to analyse numerical or experimental data. It is usually difficult to obtain a pressure map experimentally, PIV measurements giving only acces to velocity fields. A natural choice to extract coherent structures from PIV measurements would be the study of the instantaneous vorticity field, whose peaks coincide with coherent structures. However, it has been shown that vorticity cannot permit to distinguish between swirling motion of a vortex from pure shearing motions \citep{Jeong:1995, Kolar:2007, Schram:2004}. This is why several methodologies based on the analysis of the velocity gradient tensor have been developed during the last two decades. A recent review of these methods is proposed in \citep{Kolar:2007}, among others.\\
One of the most widely used criterion to extract coherent structures from PIV measurements, known as \textit{the $\lambda$2 criterion}, has been proposed by \citet{Jeong:1995}. They define a vortex in terms of eigenvalues of the symmetric tensor $\tens{S}^2 + \tens{\Omega}^2$ where $\tens{S}$ and $\tens{\Omega}$ are the symmetric and antisymmetric parts of the velocity gradient tensor $\nabla \bf{u}$. As $\tens{S}^2 + \tens{\Omega}^2$ is symmetric, it has only real eigenvalues ordered as follows $\lambda1 > \lambda 2 > \lambda 3$. According to this criterion, a local minimum of the $\lambda 2$ field is associated with the local pressure minimum that is met in the vortex core. The study of the $\lambda 2$ field permits to discriminate between swirling and shearing motions. However the calculation of the $\lambda 2$ field requires the velocity field and its variation in the three directions $(\vec{x}, \vec{y}, \vec{z})$ which are not trivial to obtain experimentally. With 2D-PIV, only the two components of the velocity fields contained in the laser sheet plane and their variations in this plane are available. Because of this limitation and assuming that the flow is locally two-dimensional, which is a strong assumption here, the simplification of the $\lambda 2$ criterion can be used \citep{Schram:2004, Anthoine:2003}:
\begin{equation}
\lambda 2 =\left( \frac{\partial{u}}{\partial{x}}\right)^{2} + \left( \frac{\partial{v}}{\partial{x}}\right)\left( \frac{\partial{u}}{\partial{y}}\right)
\end{equation}
This criterion indicates that within the vortex, the derivatives of the velocity components must be of opposite signs. Thus, the diameter $D_v$ of the vortex is defined by the zero crossing contour of $\lambda 2$. Finally, it has to be noted that the $\lambda 2$ field is a Galilean invariant, so that no additional transport should affect it.\\
The instantaneous $\lambda 2$ fields are first computed from the instantaneous velocity vector fields and then processed to extract vortices. The detection method is based on the selectivity property in space and scale of the wavelet transform of the instantaneous $\lambda 2$ field \citep{Schram:2004, Anthoine:2003}. Assuming that the vortex signature in a $\lambda 2$ field is a Gaussian curve, this allows the determination of the vortex position and size. To detect such a shape, \textit{the Maars mother wavelet} has been selected to perform the wavelet analysis. The mother wavelet is given by:
\begin{equation}
\Psi(x,y) = (2-x^{2}-y^{2})\textrm{ exp }\left(-\frac{x^{2}+y^{2}}{2}\right)
 \end{equation}
The wavelet transform of the $\lambda 2$ field is obtained by the convolution product:
 \begin{equation}
W = \big<\Psi_{l,x',y'}\big|\lambda 2\big> = \int^{+\infty}_{-\infty} \int^{+\infty}_{-\infty} \lambda 2 (x,y)\Psi_{l,x',y'}dxdy
\end{equation}
where $l$ is the scale of the mother wavelet, and $x'$ and $y'$ are defined by: $x'\, =\, x/l$ and $y'\, =\, y/l$. The detection algorithm first permits to calculate the wavelet coefficient, $|W|$, on each location $(\vec{x}, \vec{y})$ of the PIV plane. Then a vortex is identified if $|W|$ is higher than a chosen threshold value and if it does not overlap another vortex.
The detection algorithm provides the center of the vortices that are detected, assuming that each vortex center coincides with a local maximum of $|W|$. This makes it possible to estimate the velocity distribution around the vortices and their convection speed. The convection (or transport) velocity of a structure ($u_{c}; v_{c}$) is determined by averaging the velocity vectors contained in the structure, while the velocity distribution in the vortex core is obtained by substracting the convection velocity vectors to the instantenous field.
These informations are necessary to follow vortex displacements in the combustion chamber and help in determining their appearance frequency in the chamber.
More details about the wavelet analysis can be found in refs  \citep{Schram:2004, Anthoine:2003}.

\section{Experimental results and interpretations} \label{sec:55}
\subsection{Operating conditions} \label{sec:5}
\begin{figure}[!h] 
\centering
\includegraphics[width=4cm]{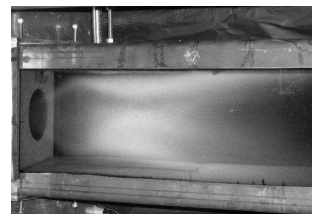} 
\caption{Natural emission of the flame. $P = 74$~kW, $Q_{air} = 105$~Nm$^{3}$.h$^{-1}$, $Q_{C3H8} = 2.95$~Nm$^{3}$.h$^{-1}$ and $\Phi$ = 0.67.}
\label{fig:4}
\end{figure}

\begin{figure}[!h] 
\centering
\includegraphics[width=8cm]{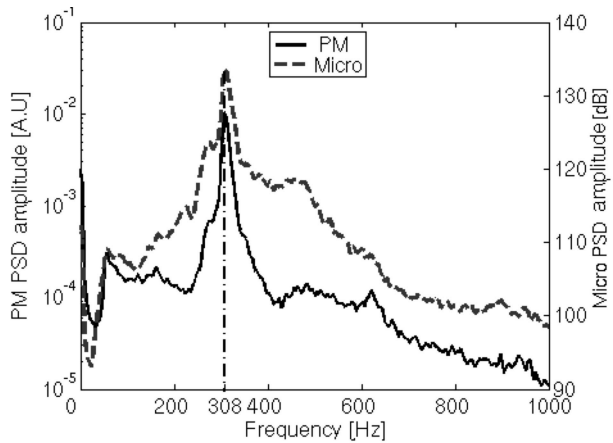}
\caption{PSD amplitude of microphone (dashed line) [dB] and PMT (Solid line) [logarithmic scale]. $P = 74$~kW, $Q_{air} = 105$~Nm$^{3}$.h$^{-1}$, $Q_{C3H8} = 2.95$~Nm$^{3}$.h$^{-1}$ and $\Phi$ = 0.67.} 
\label{fig:5}
\end{figure}

\begin{figure}[!h] 
\centering
\includegraphics[width=8cm]{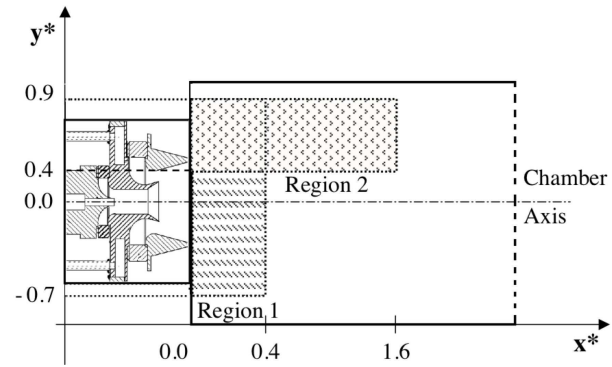} 
\caption{Scheme of the two regions defined for the PIV measurements, with $x* = x/D$ and $y*=y/D$. $x^* = 0$ corresponds to the begining of the combustion chamber, $y^* = 0$ coincides with the axis of the combustion chamber.}
\label{fig:6}
\end{figure}

The HFPIV diagnostic is tested for the operating conditions reported in table~\ref{tab:2}.
\begin{table}[!h]
\centering
\caption{Operating Conditions. $\bf{(\textrm{Q}_{C3H8})_{prim}}$ and $\bf{(\textrm{Q}_{C3H8})_{sec}}$ are the fuel flow rates injected respectively in the primary and the secondary stages, while $\bf{\textrm{Q}_{air}}$ is the global air flow rate. $\Phi$ is the global equivalence ratio and P is the power delivered by the burner.}
\label{tab:2}       
\begin{tabular}{lllll}
\hline\noalign{\smallskip}
Q$_{air}$&(Q$_{C3H8}$)$_{prim}$&(Q$_{C3H8}$)$_{sec}$&P&$\Phi$\\ 
\noalign{\smallskip}\hline\noalign{\smallskip}
Nm$^{3}$.h$^{-1}$ &Nm$^{3}$.h$^{-1}$&Nm$^{3}$.h$^{-1}$& kW & {} \\ 
\noalign{\smallskip}\hline\noalign{\smallskip}
105 & 0 & 2.95 & 74 & 0.67 \\
\noalign{\smallskip}\hline
\end{tabular}
\end{table}\\
To ease the analysis proposed in the present paper, only the second stage of the injector has been fed with propane. A complete analysis of the influence of staging on the burner stability will be proposed in a near future.\\
The power associated with this regime is 74~kW. It may appear low compared to industrial applications, but is already quite high for a laboratory scale burner. A picture of the flame is shown on figure~\ref{fig:4} for this particular regime.
\\
An experimental study of the acoustic behavior of the burner has shown that in the operating conditions chosen here, the burner exhibits strong combustion instabilities. This study has been carried out using a calibrated 1/4" microphone and a photomultiplier tube (PMT) equipped with a filter centered on the chemiluminescence band of $CH$* free radical (416 nm) \citep{Higgins:2001}. The microphone (\textit{Bruel} \& \textit{Kjaer} 4136) placed in the bottom wall of the chamber  measures the pressure fluctuations ($p$') on the semi-length of the chamber, while the PMT (\textit{Electron-Tubes}, type 9124QB) collects the ligth emitted by the whole flame. The PSD of $p$' and the fluctuating part of $CH$* emissions ($q$') are reported on figure~\ref{fig:5}. Both PSD amplitudes present a strong peak, which corresponds to a frequency $f = 316 \textrm{ Hz}$, meaning that thermo-acoustic instabilities are probably observed.
\\
This frequency can be associated with the quarter wave mode of the combustion chamber. Indeed assuming that this chamber is homogeneously filled with burnt gases of average temperature $T_{b}$, the acoustic modes of the combustion chamber can be estimated as: $f = (2n+1)c_{b}/(4L)$, where $c_{b}$ is the sound velocity corresponding to $T_b$ ($c_{b}=\sqrt{\gamma r T_{b}}$), $L$ is the combustion chamber length and $n$ is the order of the resonant mode. It may be assumed that the average properties of the gaseous medium are equivalent to the ones of air. If we finally assume that $1000\textrm{K}<Tb<1500\textrm{K}$, the quarter wave mode ( defined by $n = 0$) can be associated with frequencies between 300 and $375\textrm{ Hz}$, a range which is consistent with the experimental results.
It is expected from PIV measurements to gain insight into the dynamical coupling leading to combustion instabilities.\\
Once the flow reaches a steady state and the flame is well stabilized in the combustion chamber, PIV images are acquired in the combustion chamber downstream the injection plane.
Two complementary regions are defined for the measurements, as visible on figure~\ref{fig:6}, near the injection plane. The notations $x^*$ and $y^*$ are defined by  $x^* = x/D$ and $y^* = y/D$. The first region corresponds to an observation area which is 512 pixel high and 128 pixel wide and permits to visualize the flow field in a vertical window ($0 < x^*< 0.4$ and $-0.7 < y^*< 0.9$). The second region is a 128 pixel high and 512 pixel wide window observation, that focuses on the phenomena that take place in the top part of the combustion chamber, between $0 < x^* < 1.8$ and $0.4 < y^* < 0.87$. While the first region makes it possible to analyze the distribution of vortices between the top and bottom parts of the chamber, the second one is devoted to the convection of structures along the chamber wall.
\\
Several validation routines have been used to perform data analysis:
\begin{itemize}
\item \textit{Peak-height validation} validates or rejects individual vectors based on the value of the peak height in the correlation plane where the vector displacement is measured. The detectability criterion \citep{Keane:1992}, $k$, that compares the highest validation peak with the second one is used to perform this validation method.
\item \textit{Velocity range validation} rejects vectors which are outside a certain expected range of velocities in the flow. This can be checked component by component or directly on the velocity norm.
\item \textit{Moving-average validation} validates or rejects vectors using a comparison between the vector tested and the average of the vectors in a rectangular neighbourhood, using the acceptance factor, $\alpha$.  This validation method is a particular case of iterative filtered validations \citep{HostMadsen:1994}.
\end{itemize} 
As a result of a parametric study of the influence of validation methods on the PIV results \citep{barbosa:2008}, the values $k = 1.2$ and $\alpha = 0.1$ seems to be the most adapted to the present measurements.\\
Nevertheless, results presented in the following sections depend on the data post-processing and particularly on the validation methods used. The parametric study has shown that changing the validation parameters value introduces a maximal uncertainty of 11~\% in the maximal velocity, while it has no effect on the position of the maximal and minimal velocity regions \citep{barbosa:2008}.

\subsection{Mean flow field} \label{sec:6}
\begin{figure}[!h] 
\centering
\includegraphics[width=8cm]{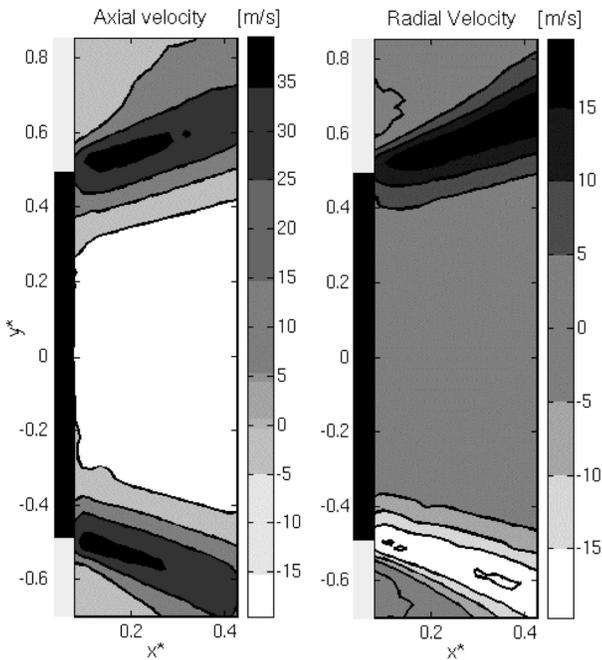} 
\caption{Contours of the mean axial velocity $U_{mean}$ and mean radial velocity $V_{mean}$. The velocity field is measured on the axis of the combustion chamber. The edges of the chamber and the injection area are drawn respectively in black and in gray. }
\label{fig:7}
\end{figure}

\begin{figure}[!h] 
\centering
\includegraphics[width= 8cm]{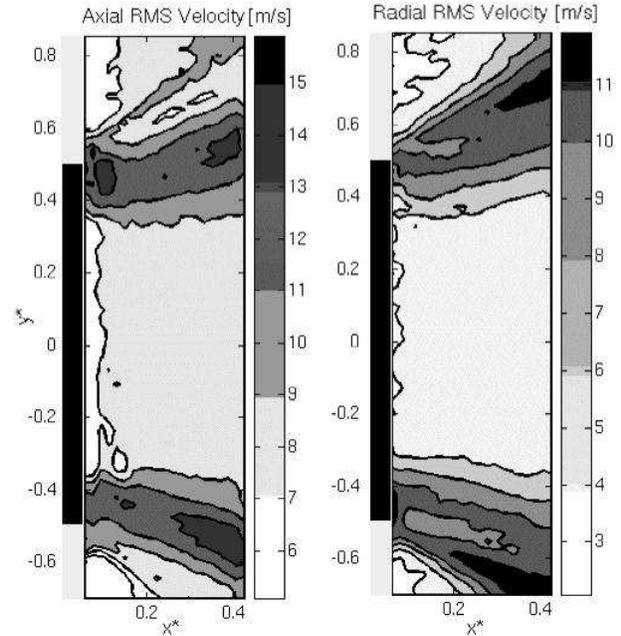} 
\caption{Contours of the axial RMS velocity $U_{rms}$ and radial RMS velocity $V_{rms}$. The edges of the chamber and the injection area are respectively drawn in black and in gray.}
\label{fig:8}
\end{figure}

\begin{figure}[!h] 
\includegraphics[width=8.4cm]{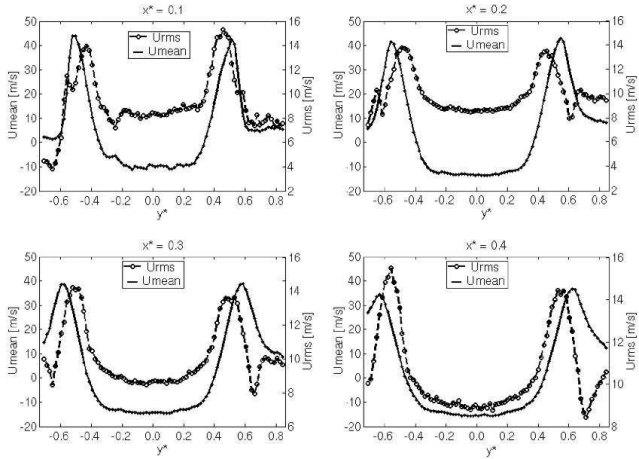} 
\caption{Profiles of  the mean axial velocity $U_{mean}$ and the axial RMS velocity $U_{rms}$  for four successive vertical cuts, $x^* = 0.1$, $x^* = 0.2$, $x^* = 0.3$ and $x^* = 0.4$.}
\label{fig:9}
\end{figure}

\begin{figure}[!h] 
\centering
\includegraphics[width=8.4cm]{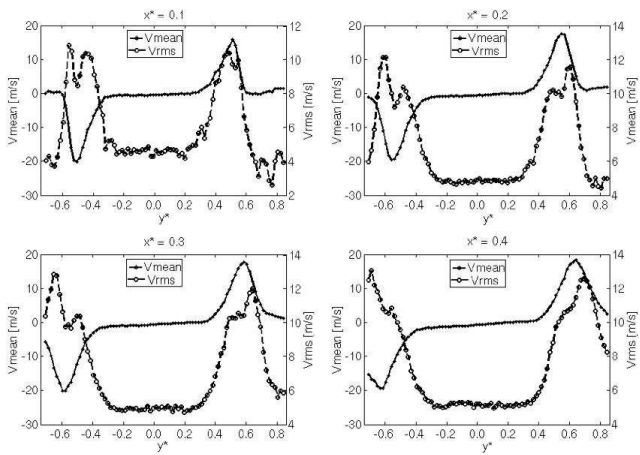} 
\caption{Profiles of the mean radial velocity $V_{mean}$ and the RMS of radial velocity $V_{rms}$ for four successive vertical cuts, $x^* = 0.1$, $x^* = 0.2$, $x^* = 0.3$ and $x^* = 0.4$.}
\label{fig:10}
\end{figure}

\begin{figure}[!h] 
\centering
\includegraphics[width=8cm]{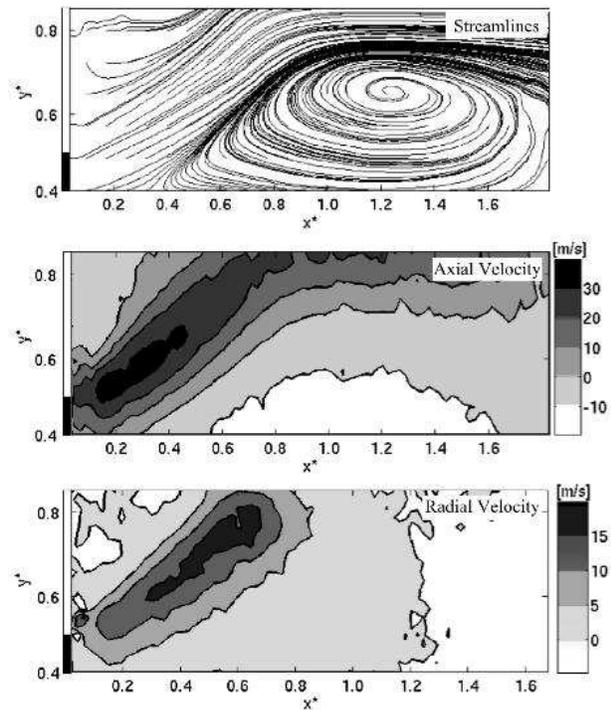} 
\caption{Streamlines of the mean velocity field are plotted on the top part of the figure. Only the top part of the combustion chamber is represented. The mean axial and radial velocity are displayed in the middle and the bottom parts of the figure. The velocity field is measured on the axis of the combustion chamber.}
\label{fig:11}
\end{figure}

\begin{figure}[!h] 
\centering
\includegraphics[width=8.4cm]{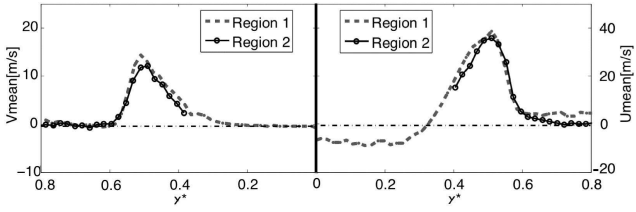} 
\caption{Profiles of $V_{mean}$ (rigth) and $U_{mean}$ (left) for $x^* = 0.1 $. Profiles obtained with the vertical observation area ("region 1") are plotted in gray and the ones obtained with the horizontal observation area ("region 2") are plotted in black.}
\label{fig:12}
\end{figure}

\begin{figure}[!h] 
\centering
\includegraphics[width=7cm]{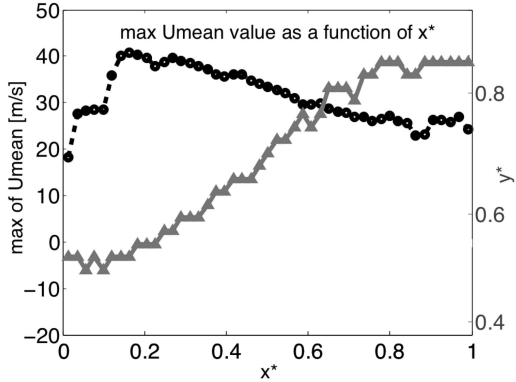} 
\caption{Evolution of the maximum value of $U_{mean}$ with $x^*$. The evolution of the normalized radial position corresponding to this maximum $U_{mean}$ value is also reported. }
\label{fig:13}
\end{figure}

\begin{figure}[!h] 
\centering
\includegraphics[width=7cm]{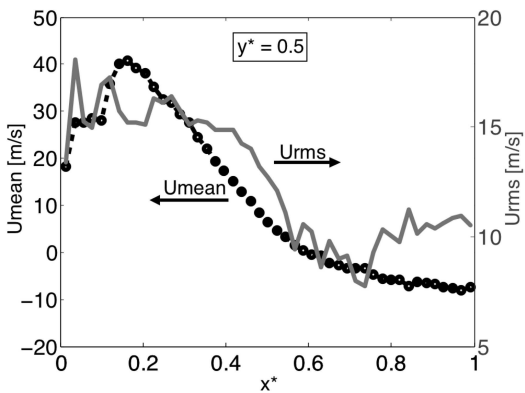} 
\caption{Evolution of $U_{mean}$ and $U_{rms}$ with $x^*$ for $y^* \, =\, 0.5.$ }
\label{fig:14}
\end{figure}

The mean axial and radial velocities, measured in a plane placed on the chamber axis, are reported on figure~\ref{fig:7}. The top of this vertical observation area is at the distance $y^*= 0.87$, while the top (resp. the bottom) of the chamber is at a distance of $y^* =1$ (resp. $y^* = -1$) and the edges of the injector are situated at $y^* = 0.5\, (\textrm{resp. }-0.5)$.
\\
First, a large negative velocity area near the burner axis is observed on figure~\ref{fig:7}. It corresponds to the central recirculation region, which provides the major mechanism for flame stabilization in a swirl-stabilized burner.  Second, the flow exhibits the features of an annular conical jet whose internal boundary is visible on both sides of the cut at the distances $y^*=0.4$ and $-0.4$. This observation indicates that the flame is clearly stabilized within the outer divergent of the injector. The maxima of the mean axial and radial velocities are observed within this annular jet. It can be noted that the maximum level of the radial mean velocity is almost one third of the axial component value. Two recirculation areas can also be observed in the top and the bottom corners of the chamber due to the sudden expansion at the injector exit plane.
The annular jet induces two shear layers, which coincide with the locations of the regions of maximum RMS velocity components, as visible on figure~\ref{fig:8}. As expected, the lower levels of turbulent intensities are obtained in the inner and outer recirculation areas.
\\
The distribution of the mean and RMS velocities are reported on figure~\ref{fig:9} (axial component) and figure~\ref{fig:10} (radial component) for four successive axial cuts from $x^* = 0.1$ to $x^*=0.4$.\\
The $U_{mean}$ profile, plotted in figure~\ref{fig:9}, is symmetric around $y^* = 0$. At $x^* = 0.1$, it presents two peaks at a distance $y^* = \pm 0.5$, corresponding to the annular jet described before. These data will be used in section~\ref{sec:7} to calculate the associated massflow rate and compare it with the value given by the massflow controller.
\\
A large negative velocity area, corresponding to the inner backflow region, extends between $y^*= -0.35$ and $y^*=0.35$.  Increasing $x^*$, the location of the maximum level of $U_{mean}$ moves away from the axis of the injector, while the maximum of $U_{mean}$ decreases from an initial value of 44~m.s$^{-1}$ ($x^* = 0.1$) to a final value of 37~m.s$^{-1}$ when $x^* = 0.4$. Figure~\ref{fig:9} also permits to compare the distributions of $U_{mean}$ and $U_{rms}$. Profiles of $U_{rms}$, plotted in figure~\ref{fig:9}, are symmetric around $y^* = 0$. $U_{rms}$ presents maxima on each side of the chamber at $y^*\approx \pm0.43$, i.e where the mixing region between the central recirculation zone and the positive streamwise velocity flow is located. Two secondary peaks, which are associated with a far lower value and reached at $y^*\approx \pm0.55$, indicate the location of the external shear layer. When $x^* >  0.2$, the $U_{rms}$ profile presents only one peak in the mixing region between the recirculation area and the annular jet. Indeed, the outer recirculation has almost disappeared when $x^* > 0.2$ (figure~\ref{fig:7}).\\
The same phenomena are visible on figure~\ref{fig:10}, where the distributions of radial velocity $V_{mean}$ and $V_{rms}$ are reported. When $x^*< 0.2$, the profile of $V_{rms}$ presents two peaks distributed on both sides of the annular jet as in the cse of $U_{rms}$. However, when $x^* > 0.2$, $U_{rms}$ and $V_{rms}$ do not follow similar trends anymore: the magnitude of $V_{rms}$ increases monotonicly with the radial distance ($y^*> 0$), so that the peak of $V_{rms}$ is observed near the top (resp. the bottom) of the combustion chamber. Due to the coarse spatial resolution of the measurement technique, this point needs to be further investigated.
\\
In order to better visualize the phenomena occurring around the jet, the camera focuses on the top of the chamber ($0< x^*< 1.8$ and $0.4< y^*< 0.87$). Figure~\ref{fig:11} displays the mean velocity field obtained by processing $N = 12 000$ instantaneous fields. The central back-flow region, revealed by the streamline plot and the negative axial velocity area, extends between $x^* =0.4$ and 1.8, while the second recirculation area, in the top of the chamber ($y^* > 0.6$) develops between $x^* = 0$ and 0.3. The annular jet characterized by positive velocity vectors separes these two recirculation areas.
\\
The distributions of the mean axial and radial velocities obtained with a vertical  observation area ("Region~1") and a horizontal one ("Region~2") are plotted on figure~\ref{fig:12} for the vertical cut $x^* = 0.1$. The two measurements show a very good agreement in terms of $U_{mean}$ and $V_{mean}$. The observation area has no effect on the position of the velocity peaks, while a difference in amplitude of less than 5~\% is observed in the region of high velocities, which is quite acceptable.
\\
Using this recording, the evolution of the maximum of $U_{mean}$ in each vertical cut is plotted in figure~\ref{fig:13} as a function of the reduced axial distance to the injection plane. On the same graph, the evolution of  the radial position of this maximum velocity is reported using the left ordinate axis.
\\
The $U_{mean}$ peak level increases initially up to a distance of $x^* = 0.2$ where it reaches 43~m.s$^{-1}$ for a transverse distance $y^* = 0.5$. Further downstream the $U_{mean}$ level decreases from $43 \, \textrm{m.s}^{-1}$ to $25 \, \textrm{m.s}^{-1}$, while the radial position of the peak velocity increases from $y^* = 0.5$ to 0.8. It can be considered as indicative of the position of the flame front in the chamber, assuming that the positive streamwise velocity separates the flame from the fresh gases.\\
The distributions of $U_{mean}$ and $U_{rms}$ are finally plotted  on figure~\ref{fig:14} in the shear layer vicinity ($y^* = 0.5$).
\\
As already mentioned, $U_{mean}$ reaches its maximum level at $x^* = 0.2$. The profile of $U_{rms}$ downstream the injection plane present two peaks at $x^* = 0.1$ and 0.25, corresponding to the location of the external and internal shear layers due to the annular  conical jet. Close to the wall, $U_{rms}$ presents a peak, which seems to have no physical signification and might be due to the boundary effects. The profile of $U_{rms}$ indicates that the maximum of fluctuations takes place downstream the injection plane: $x^*< 0.6$. Farther downstream ($x^*> 0.6$), the mean axial velocity becomes negative and $U_{rms}$ reaches its lowest level. This indicates that the internal recirculation area extends almost beyond 2 diameters from the burner.
\subsection{Calculation of the flow rate injected} \label{sec:7}				
In order to check the coherence of the experimental results, the fluid flow rate in the combustion chamber is calculated using the axial velocity measured experimentally using HFPIV. The fluid flow rate is estimated using the radial distribution of Umean, taken at a distance of $x^* =  0.06$, so that the result is less disturbed by edge effects. Assuming that the flow is symmetric, the flow rate is calculated using the following definition:		
\begin{equation}
Q = \int{\int_{S}{ \bf{v}.\bf{n}} dS}
\end{equation}
where $\bf{v}$ is the velocity vector. Imposed and measured results are respectively $Q_{th} = 113$~m$^3$.h$^{-1}$ and $Q_{exp}=115$ m$^3$.h$^{-1}$, meaning that the relative error between imposed and measured flow rates is about 2\% for the case studied. 
This difference can be due to the precision of the mass flow meters, the precision of the PIV measurements, the assumption of axisymmetry of the flow or the position of the profile used. However the relative error is weak, and the experimental results seem consistent. Finally, this good agreement can be used to say that PIV measurements give a good description of the phenomena taking place in the combustion chamber on a quantitative point of view.
The mean flow field structures have been characterized using PIV measurements. However, the principal advantage of this diagnostic is the high acquisition rate of the image pairs, which are then processed to calculate instantaneous velocity fields. An analysis of the instantaneous flow field evolution in the chamber is reported the following section. 
\section{Analysis of instantaneous flow fields} \label{sec:8}
\begin{figure*} 
\centering
\includegraphics[width=14.5cm]{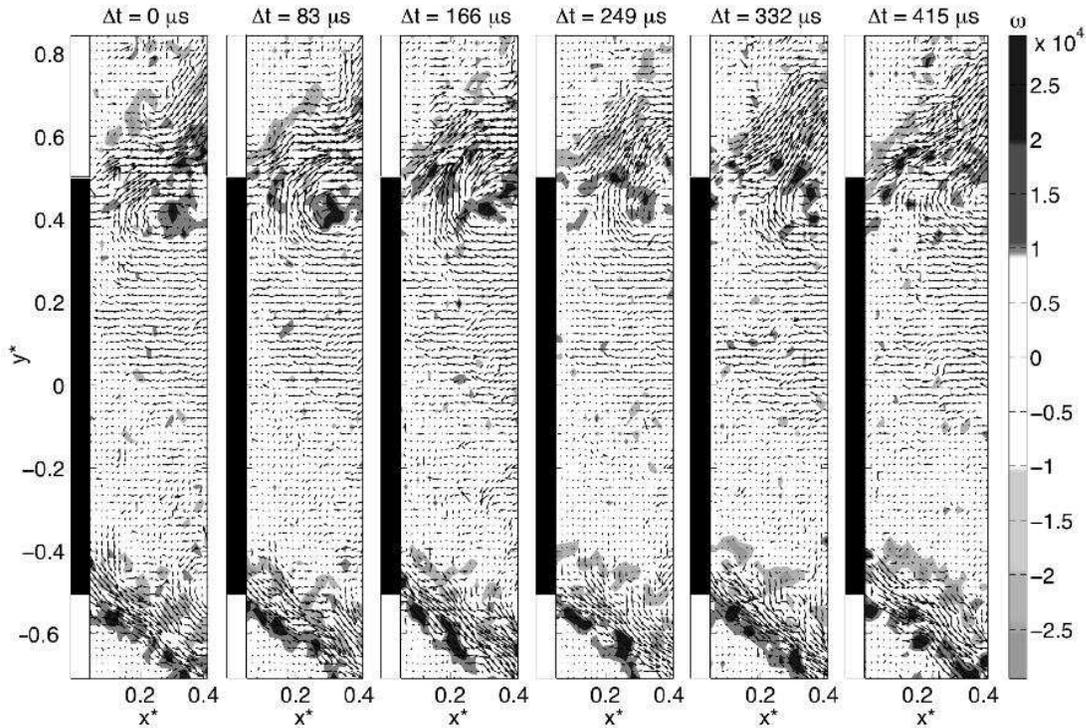} 
\caption{Instantaneous velocity vectors and associated vorticity fields for six successive recordings separated by a time delay of $83\, \mu$s. These fields are obtained on the axis of the combustion chamber. On each picture, the location of the injector outlet is represented in black. }
\label{fig:15}
\end{figure*}

A temporal analysis of the velocity field is carried out in the regime defined in table~\ref{tab:2}. A serie of six successive instantaneous velocity vectors and their associated vorticity field are displayed in figure~\ref{fig:15}. The views are taken on the vertical symmetry axis of the combustion chamber ($y^* = 0$). A time delay of $\Delta t = 83\, \mu$s separates each recording.
This serie has been chosen because it is representative of the phenomena observed.
\\
The instantaneous velocity fields exhibit large deviations from the mean velocity field described on figures~\ref{fig:7} and~\ref{fig:8}. Still, a large inner recirculation area and an annular conical jet are clearly visible on each picture. The inner recirculation area, characterized by a negative velocity and a low vorticity level, extends on the center of each picture. The location of the boundary of the annular jet changes dramatically from one instantaneous image to the others. Figure~\ref{fig:15} shows that maximum levels of vorticity, $\omega$, coincide with the jet boundaries, where the flow is characterized by a great shear rate. Finally, a large scale vortex is visible at $y^* \approx 0.4$ on the first frame of figure~\ref{fig:15}. The five other frames show the deformation and convection of this structure, and a modification of its vorticity value.\\
As seen before a straightforward criterion to identify a vortex is to detect the presence of a vorticity peak. However, while the center of a vortex is actually associated with a vorticity maximum, a vorticity peak does not necessary correspond to a vortex center. Indeed, on figure~\ref{fig:15}, high values are present on the external shear layer surrounding the annular jet, where no vortex can be detected. An analysis based on the $\lambda  2$~method presented by Jeong and Hussain \cite{Jeong:1995} and describe in paragraph~\ref{sec:4} is then proposed to identify vortices. 
\subsection{Vortex detection}\label{sec:9}
\begin{figure*}  [!t]
\centering
\includegraphics[width=17cm]{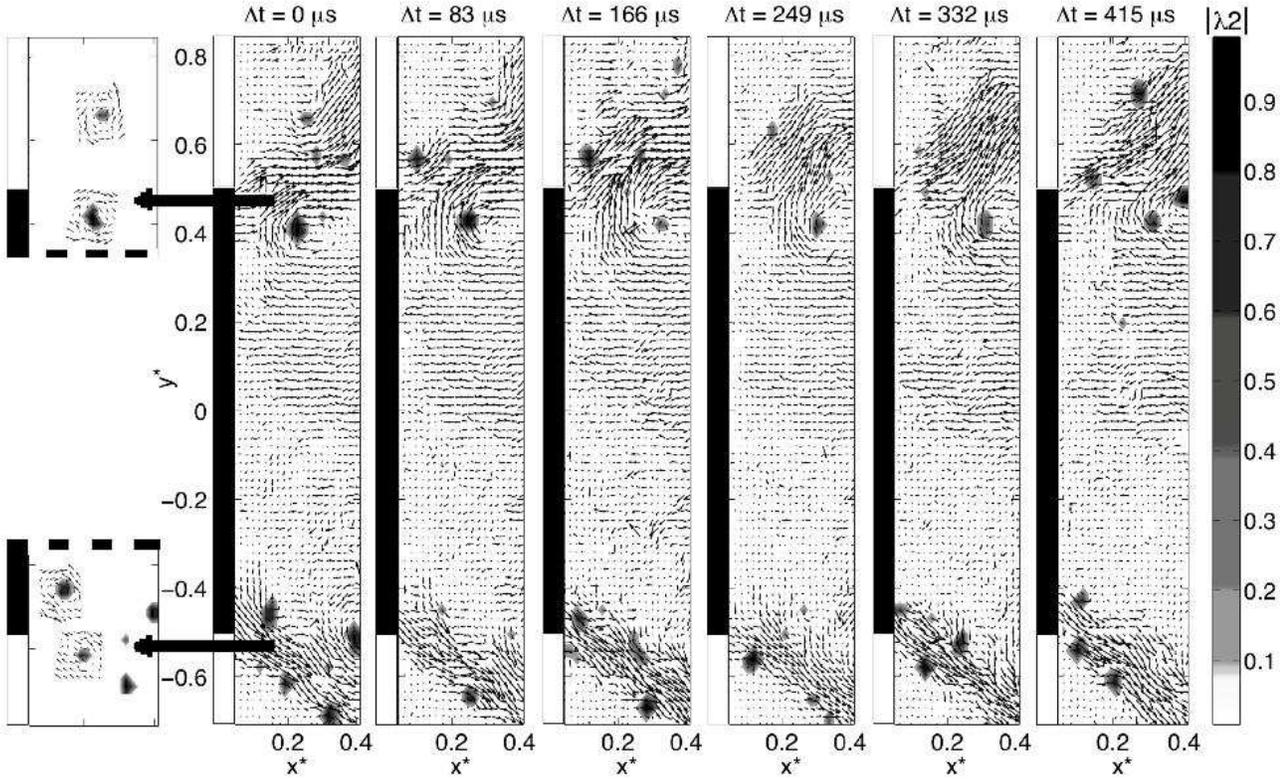} 
\caption{Velocity vectors and associated wavelet transform of the $\lambda$2 field for the six successive recordings shown on figure~\ref{fig:15}. For $\Delta t = 0\, \mu$s, the velocity field around each structure detected is shown. On each frame, the location of the injector outlet is represented in black. Left: Enlargement of two zones where vortices have been detected.}
\label{fig:16}
\end{figure*}

\begin{figure}[h!] 
\centering
\includegraphics[width=8cm]{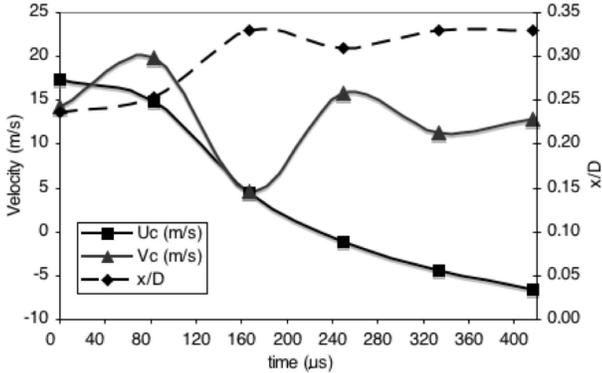} 
\caption{Distribution of the axial position, the axial and radial transport velocities as a function of time, for the six fields of figure~\ref{fig:16}. The radial position of the followed structure is  $y^* \approx 0.4$. The time $\textrm{t = 0 }\mu s$ coincides with the first frame of figure~\ref{fig:16}.}
\label{fig:17}
\end{figure}

Figure~\ref{fig:16} shows the $\lambda$2 fields, analyzed with the wavelet  transform, for the six velocity fields of figure~\ref{fig:15}. The $\lambda 2$ criterion is not sensitive to shear regions, on the contrary to vorticity. Several peaks of $\lambda$2 are detected and the structure cited in the previous paragraph (figure~\ref{fig:15}) is visible in the top part of the combustion chamber ($y^* \approx 0.4$).
\\
Using the detection algorithm described in paragraph~\ref{sec:4}, the axial displacement and the convection velocity of this structure are determined and their evolution as a function of time is reported on figure~\ref{fig:17}. 
\\
First, for $t < 160\, \mu$s, the axial convection speed $U_{c}$ is large and positive, explaining the displacement of the structure towards the right end of the observation area. When, $t > 160\, \mu$s, the value of $U_{c}$ becomes slightly negative explaining that the location of the structure does not change, as if the structure was trapped at the limit of a region where the flow recirculates. The value of the radial transport velocity, $V_{c}$, is almost constant, close to 12~m.s$^{-1}$.\\
In the 12000 instantaneous velocity fields of the test, the phenomena described on figures~\ref{fig:15},~\ref{fig:16} and~\ref{fig:17} have been periodically observed on both half parts of the chamber. The values obtained on figure~\ref{fig:17} give a good estimate of the convection speed of structures created within the combustion chamber. Lastly, as $\lambda 2$ is a Galilean invariant, the detection algorithm permits to find small vortices that are difficult to visualise due to the high levels of velocity encoutered in the chamber as shown on the details of the first frame of figure~\ref{fig:16}. In spite of the coarse grid used to calculate the velocity vector field, these structures are detected.

\subsection{Coherent structure frequency} \label{sec:10}
\begin{figure}  [h!] 
\centering
\includegraphics[width=8cm]{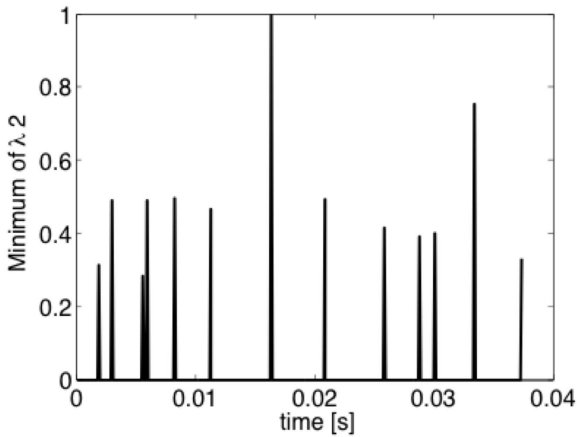} 
\caption{Vortices detected within a squared observation window 3 mm wide centred on the point of coordinates ($x^*=0.24$, $y^*=0.42$).}
\label{fig:18}
\end{figure}

\begin{figure*}  
\centering
\includegraphics[width=17cm]{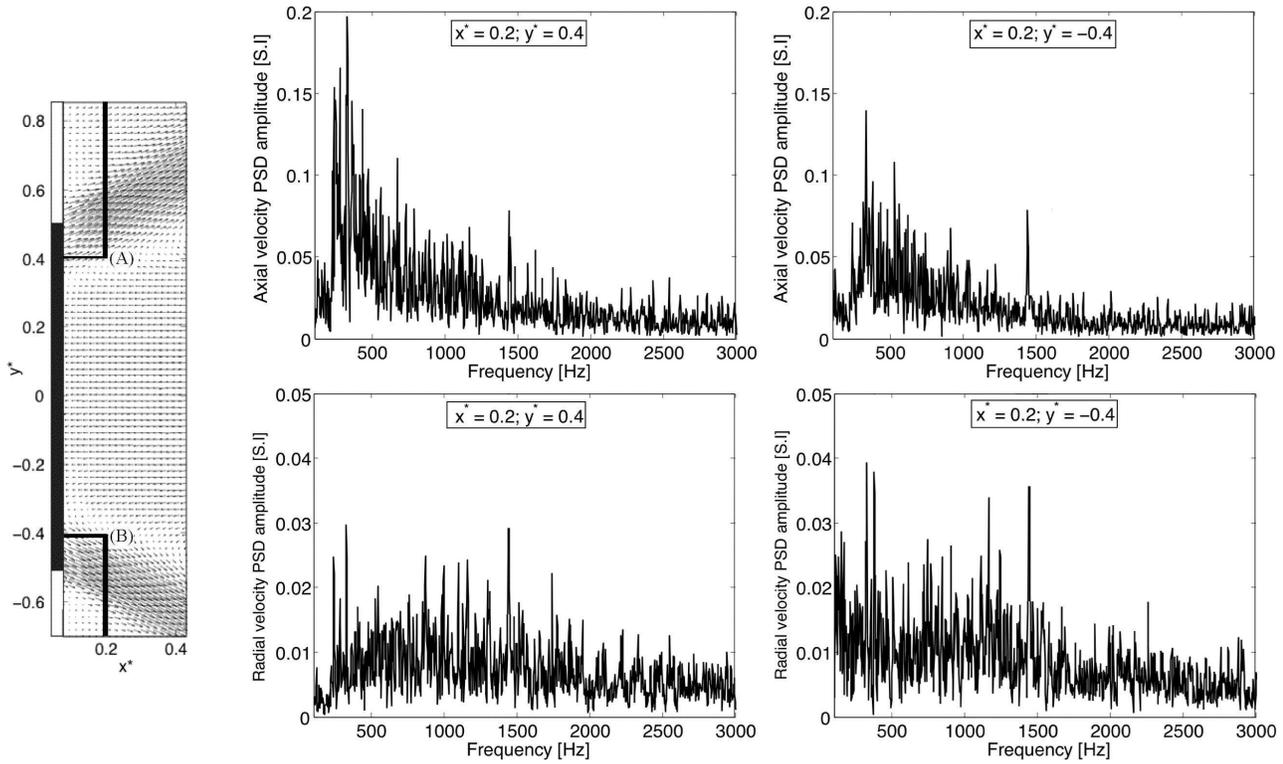} 
\caption{PSD amplitude of the axial and radial velocities at point $A$ of coordinates ($x^*=0.2$, $y^*=0.4$) and point $B$ of coordinates ($x^*=0.2$, $y^*=-0.4$).}
\label{fig:19}
\end{figure*}

The coherent structure frequency is another important parameter for the burner characterization. As the detection algorithm gives the coordinates of each detected vortex, it is possible to count the number of vortices that crosses a given interrogation area of the chamber during a given  time (1~second for the present study). For a given observation area, if the value of $|\lambda 2|$ is higher than a given threshold a vortex is detected. The number of the frame and the value of $|\lambda 2|$ is noted so that it is possible to plot the peak of $|\lambda 2|$ versus time as visible on figure~\ref{fig:18}. In order to illustrate phenomena taking place in the chamber, a squared observation area of 3~mm wide is centred on the point of coordinates ($x^* = 0.24$, $y^* = 0.42$).
\\
On figure~\ref{fig:18}, all peaks of $|\lambda 2|$ have been normalized using the maximum value encountered during 1~second of recording.
The analysis of this interrogation area indicates a vortex frequency of about 310~Hz. The same analysis has been performed using several observation area located on the top or the bottom part of the annular jet and the frequency of vortices has always been found between 300 and 320~Hz, which corresponds to an uncertainty of about 3~\%.
To complete this analysis, the power spectral density (PSD) of the instantaneous axial and radial velocities are calculated in each point of the velocity grid. For each experiment, the acquisition frequency is $f = 12\, 000$~Hz and the number of samples is $N = 12\, 000$, so that the spectral resolution is $\Delta f = 1$~Hz.\\
To illustrate the results of the burner aerodynamic spectral analysis, the velocity PSD amplitudes calculated at the point $A$ ($x^* = 0.2; \, y^* = 0.4$) and the point $B$ ($x^* = 0.2;\, y^* = -0.4$) are plotted versus frequency in figure~\ref{fig:19}. The points $A$ and $B$ are located at the boundary between the central re-circulation area and the annular jet. As seen before vortices are mainly created in this region of great shear rate. These two points are chosen symmetric around the chamber axis.
\\
The PSD amplitude of $u$ and $v$ calculated at point $A$ are reported on the left part of figure~\ref{fig:19} and the same analysis has been performed in the bottom part of the combustion chamber, for point $B$. The PSD amplitudes of the axial and radial velocity calculated are reported on the right part of figure~\ref{fig:19}. The PSD amplitude of $u$ indicates that the burner behaves like a low-pass filter: the highest levels of the PSD amplitude are observed in the range of frequencies $0<f<1000$~Hz and then the PSD amplitude decreases to reach its lowest level when $f > 2000$~Hz.
The $u$ spectra for $A$ and $B$ are quite similar and they both present a peak at a frequency $f_u \approx 310$~Hz.\\
The main peak value of the $v$ spectrum is also observed at a frequency $f_v \approx 310$~Hz for point~$B$, even if this peak is less well marked. This last point is even more pronounced for the $v$ spectrum of point~$A$, where the signal is broadband in the range [0; 1000]~Hz\\
It can be noted that $f_u$ and $f_v$ are very close to the vortex frequency determined before and to the acoustic quater wave mode of the combustion chamber (paragraph~\ref{sec:5}). One can imagine a straightforward scenario, in which the flame heat release rate is pulsated by strong aerodynamic structures, whose frequency is the first eigenmode of the combustion chamber. 

\section{Concluding Remarks} \label{sec:11}
The aim of the present study was to detail and evaluate the efficiency of a system of HFPIV operating at 12 kHz. This diagnostic is used to characterise the behavior of an experimental lean premixed swirl-stabilized burner representative of a gas turbine combustor, which may exhibit strong combustion instabilities under certain operating conditions.  It is expected from HFPIV to provide well temporally resolved velocity fields that permit to better understand these complex turbulent reactive flows.\\
Using HFPIV measurements, the mean structure of the flow field has been first determined. This mean velocity field exhibits the features of a swirling flow, with an annular conical jet where high values of the axial velocity component are observed, and a large central recirculation zone that provides the major mechanism for flame stabilization. RMS velocities are also estimated, showing large levels of fluctuations inside the combution chamber, associated with a strong dynamical behavior of the flame especially in the shear regions. These mean quantities are also used to estimate the efficiency of the HFPIV system. The mass flow rate entering the chamber is calculated and compared with the value given by the mass flow controllers and a very good precision is obtained (error around 5~\%).\\
Instantaneous velocity fields are analyzed in order to better understand the interactions between vortices and the turbulent flame in the combustion chamber. Large coherent structures are periodically detected within the combustion chamber. Their convection speed and their frequency are determined and it is shown that, in resonant situations, this frequency can be related to an acoustic resonant mode of the combustion chamber.\\
To conclude, to achieve temporally resolved measurements, it has been necessary to find a compromise between spatial and temporal resolutions, due to the technical limitations of high acquisition rate cameras at the present time. The study shows that the quite low spatial resolution is high enough to characterize the flow and to detect coherent structures. Finally, the present work indicates that the high temporal resolution chosen (12~kHz) is necessary to study turbulent GT flames where strongly unsteady phenomena occur. The actual technical developments of cameras lead to suppose that more interesting compromises between image size and acquisition rate can be found in the next years, so that it will become possible to obtain temporally resolved velocity fields measurements with HFPIV, with a better resolution. 

\begin{acknowledgements}
S\'everine Barbosa benefits from a PhD grant from D\'el\'egation G\'en\'erale pour l'Armement (DGA) and the authors wish to thank {SNECMA-SAFRAN} for the authorization to publish the present paper. The results are issued from studies supported by {SNECMA-SAFRAN} in the framework of the INCA (INitiative en Combustion Avanc\'ee) dedicated to advanced research in combustion technology.
\end{acknowledgements}

\bibliographystyle{abbrvnat} 
\bibliography{bibliopiv2}   


\end{document}